\documentclass[12pt,a4paper,final]{iopart}

\usepackage{iopams}
\usepackage{graphicx}

\usepackage{amsfonts,amsthm}
\usepackage{subfigure}
\usepackage{multirow}
\usepackage{graphicx}
\usepackage{dcolumn}
\usepackage{bm}

 \expandafter\let\csname equation*\endcsname\relax
  \expandafter\let\csname endequation*\endcsname\relax
\usepackage{amsmath,amssymb,amsthm,mathrsfs,amsfonts,dsfont}
\usepackage{enumerate}
\usepackage{subfigure}
\usepackage{xcolor}
\usepackage{color}
\usepackage{cases}
\usepackage[margin=1cm]{caption}
\usepackage{braket}

\eqnobysec

\begin{document}
\title{Practical round-robin differential-phase-shift quantum key distribution}
\date{\today}
{Zhen Zhang, Xiao Yuan, Zhu Cao, Xiongfeng Ma,$^{\ast}$\\
\address{Center for Quantum Information, Institute for Interdisciplinary Information Sciences, Tsinghua University, Beijing, China}
\eads{\mailto{xma@mail.tsinghua.edu.cn}}

\begin{abstract}
The security of quantum key distribution (QKD) relies on the Heisenberg uncertainty principle, with which legitimate users are able to estimate information leakage by monitoring the disturbance of the transmitted quantum signals. Normally, the disturbance is reflected as bit flip errors in the sifted key; thus, privacy amplification, which removes any leaked information from the key, generally depends on the bit error rate. Recently, a round-robin differential-phase-shift QKD protocol for which privacy amplification does not rely on the bit error rate [Nature 509, 475 (2014)] was proposed. The amount of leaked information can be bounded by the sender during the state-preparation stage and hence, is independent of the behaviour of the unreliable quantum channel.
In our work, we apply the tagging technique to the protocol and present a tight bound on the key rate and employ a decoy-state method. The effects of background noise and misalignment are taken into account under practical conditions. Our simulation results show that the protocol can tolerate channel error rates close to $50\%$ within a typical experiment setting. That is, there is a negligible restriction on the error rate in practice.
\end{abstract}
\vspace{2pc}
\maketitle

\section{Introduction}

Quantum cryptography enables secure information exchange between two remote parties, guaranteed by quantum physics. In particular, quantum key distribution (QKD) \cite{Bennett:BB84:1984,Ekert:QKD:1991} offers a means of distributing keys with security that is information-theoretically provable based on the fundamental laws of quantum physics \cite{mayers2001unconditional,Lo1999Science,SHORPRESKILL2000PRL,Koashi:Comp:09}. In a typical QKD protocol, a sender, Alice, transmits quantum signals through an untrusted channel to a receiver, Bob, who performs measurements and accumulates raw key data. Alice and Bob aim to share secure identical keys such that an adversary, Eve, cannot obtain information about the keys (up to a small failure probability).

Due to experimental imperfections or eavesdropping, some of the shared sifted keys of Alice and Bob are not identical. Such differences are caused by events known as bit-flip errors. Alice and Bob can run an error correction procedure to make the keys identical. Besides this, owing to eavesdropping, parts of the shared keys may not be secure. The amount of information of the shared keys that is leaked to Eve can be quantified by phase-flip errors \cite{Lo1999Science,SHORPRESKILL2000PRL}. The Heisenberg uncertainty principle tells us that any attempts to eavesdrop on the quantum channel would inevitably cause disturbance in the quantum signals. Alice and Bob can thus quantify, or at least obtain an upper bound on, the phase error rate by monitoring the disturbance, and remove it by performing privacy amplification. Finally, the ratio of the distributed secure key per sifted key bit is given by \cite{SHORPRESKILL2000PRL},
\begin{equation}\label{keyrateN}
R = 1-H(e_{\textrm{bit}}) - H(e_{\textrm{ph}}),
\end{equation}
where $e_{\textrm{bit}}$ and $e_{\textrm{ph}}$ are the bit and phase error rates, respectively, and $H(x)=-x\log_{2}x - (1-x)\log_{2}(1-x)$ is the binary Shannon entropy function.

In conventional QKD protocols, there exists a fundamental limitation on the error rate. Intuitively, the more disturbance that the adversary introduces (say, indicated by a higher bit error rate), the more information she can obtain. For example, in the BB84 protocol \cite{Bennett:BB84:1984}, due to its symmetries, the phase error rate can be estimated by the bit error rate $e_{\textrm{ph}}=e_{\textrm{bit}}$ \cite{SHORPRESKILL2000PRL}. In the extreme case, where the bit flip error $e_{\textrm{bit}}\ge11\%$, the final key rate,  $R = 1-2H(e_{\textrm{bit}})$, drops to $0$  according to Eq.~\eqref{keyrateN}, which means that no secure key can be achieved. Therefore, the above post-processing procedure works only for the case where the bit error rate is not larger than $11\%$. Higher error rate thresholds can be obtained by other postprocessing techniques \cite{1176619}, but upper bounds are generally believed to exist \cite{PhysRevLett.92.217903}.

Surprisingly, this is not the case for all QKD protocols. In a recently proposed seminal QKD protocol known as the round-robin differential-phase-shift (RRDPS) \cite{sasaki2014practical}, the phase error rate can be estimated with a different approach that does not depend on the bit error rate. Instead, the information Eve can acquire is directly bounded by the quantum source, regardless of how she interferes with the quantum signals. In this protocol, Alice encodes her information into the phase of a quantum signal that is in a superposition of $L$ optical modes (say, $L$ sequential pulses). Then, she sends the signal through a (unsafe) quantum channel to Bob, who randomly picks two of the $L$ modes and measures the phase difference between them to gain raw key data. Owing to the randomness of the measurement choices and the coherence of the signal, Eve can only acquire very limited information about the key. As the number of optical modes $L$ increases, the information that Eve can obtain by eavesdropping decreases \cite{sasaki2014practical}. With a sufficiently long quantum signal (large $L$), the phase error rate can be reduced down to $0$ and a secure key can be generated even if the bit error rate $e_{\mathrm{bit}}$ is close to $50\%$. Recently, many proof-of-principle experimental demonstrations of the RRDPS protocols have been presented \cite{PhysRevLett.114.180502, takesue2015experimental, wang2015experimental,  PhysRevA.93.030302}.
There are also several theoretical follow-ups that considered source flaws in the RRDPS protocol \cite{PhysRevA.92.060303} and its extensions to other QKD scenarios
\cite{PhysRevA.93.022330, 2016arXiv160808329C}.

In practical QKD systems, weak coherent pulses are often used as photon sources. In conventional QKD protocols, such as BB84, the multi-photon component from a coherent state cannot lead to any secure keys as it is vulnerable against the photon number-splitting attack \cite{Brassard2000PNS}. In the RRDPS protocol, when Alice splits the coherent state pulse into $L$ pulses, Bob can generate a secure key even with multi-photon components. For an $n$-photon input state, the phase error rate can be upper-bounded by $e_{\mathrm{ph}}^n\leq n/(L-1)$ \cite{sasaki2014practical}. With a sufficiently large $L$, the $n$-photon state can still positively contribute to the final key rate, according to Eq.~\eqref{keyrateN}.


When the phase is randomized, a weak coherent state can be treated as a statistical mixture of Fock states, where the photon number follows a Poisson distribution. In the original security analysis \cite{sasaki2014practical}, the phase error rate for a coherent state source is estimated by upper-bounding the photon number (up to a small failure probability). In this study, we apply the tagging technique, developed by Gottesman, Lo, L\"utkenhaus, and Preskill (GLLP) \cite{GLLP:2004}, to assess the phase error rates for different photon number states separately. As a result, we derive a tighter secure key rate bound by reducing the cost in privacy amplification. In addition, we adopt the decoy-state method \cite{PhysRevLett.91.057901,Lo:PRL:2005,WangDecoyPhysRevLett.94.230503}, which is widely used in regular QKD systems.


Furthermore, we build a simulation model to analyse the performance of the RRDPS protocol under a practical scenario. We show that, in a practical setting, the maximum transmission distance cannot infinitely increase, even if the phase error rate, $e_{\mathrm{ph}}$, drops to zero via the increase of the number of optical mode $L$. Intuitively, this is due to the fact that the background rate, which is assumed to be a linear function of $L$, limits the maximum transmission distance.
By simulation, we compare three security analysis methods: the one proposed by Sasaki, Yamamoto, and Koashi (SYK) \cite{sasaki2014practical}, and our new analysis with and without decoy states. The results show the performance improvement by our new analysis methods.

\section{Review of the RRDPS protocol}\label{RRDPS}
The RRDPS protocol is presented in figure \ref{RRDPS:protocol}. Let us first consider the case wherein Alice uses a single-photon state source. Then the state $\ket{\Psi}_{\mathbf{s}}$ that she prepares is in a superposition of $L$ optical modes,
\begin{equation}\label{eq:singlephoton}
\begin{aligned}
\ket{\Psi_{1}}_{\mathbf{s}} =  \frac{1}{\sqrt{L}}\sum_{k=0}^{L-1} (-1)^{s_k}\ket{k},
\end{aligned}
\end{equation}
where $s_k\in\{0,1\}$ is Alice's encoded key information and $\ket{k}$ denotes the state of the photon appearing in the $k$-th mode. Alice's $L$-bit key information, $\mathbf{s}\in\{0,1\}^L$, is encoded in the phase of each mode, $0$ or $\pi$. In this study, we use temporal modes as an example of optical modes and hence Eq.~\eqref{eq:singlephoton} forms an $L$-pulse sequence. In principle, Alice can use optical modes separated by other degrees of freedom, such as spectrum or angular momenta, where our results should be directly applied.

\begin{figure}[htbp]
\centering \resizebox{12cm}{!}{\includegraphics{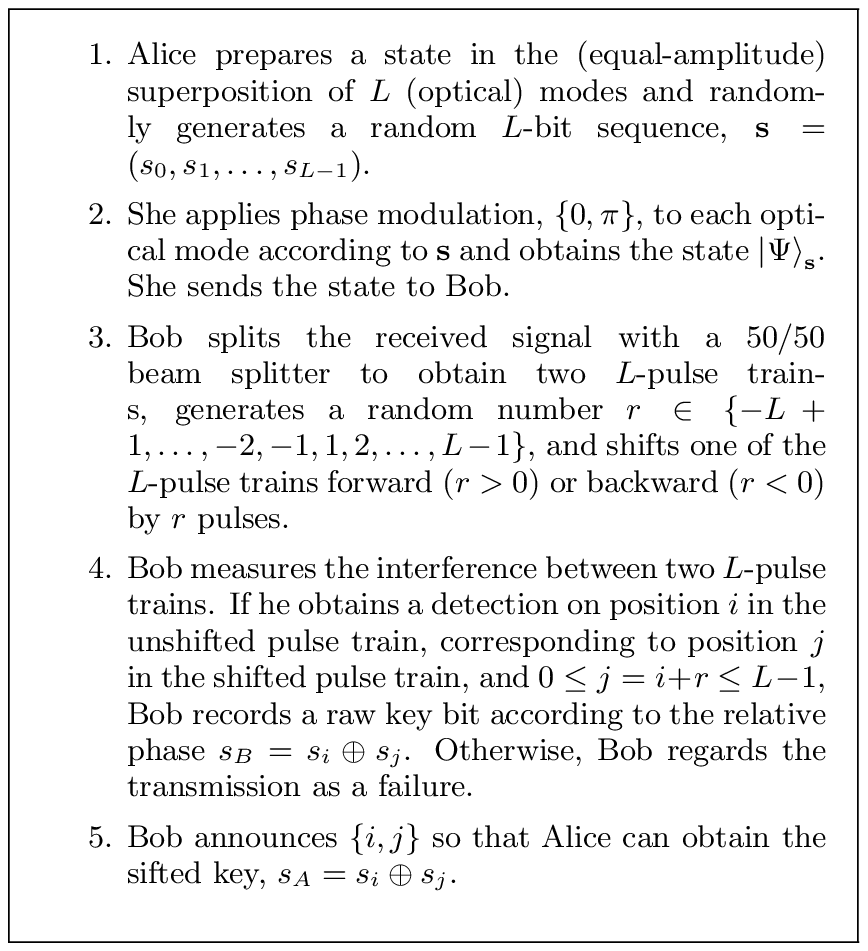}}
\caption{RRDPS protocol \cite{sasaki2014practical}.} \label{RRDPS:protocol}
\end{figure}

For general states, such as multi-photon states with extra dimensions,  Alice can also encode the key information as follows. First, Alice prepares an $L$-pulse state $\ket{\Psi}$ and $L$ ancillary qubits each in $(\ket{0} + \ket{1})/\sqrt{2}$, where $\ket{0}$ and $\ket{1}$ are the eigenstates of the $Z$-basis. Then, she applies the $L$ control operations $U = \ket{0}\bra{0}\otimes I + \ket{1}\bra{1}\otimes (-1)^{\hat{n}}$ on each of the ancillaries as control and the $L$-pulse state $\ket{\Psi}$ as the target, where $\hat{n}$ is the photon number operator. When the control qubit is $\ket{1}$, all photons in the target light are shifted by a phase $\pi$. With this, Alice finally prepares the entangled state
\begin{equation}\label{eq:State}
\begin{aligned}
\ket{\Psi}_{\mathrm{Alice}} &= 2^{-L/2}  \prod_{k=0}^{L-1} \left(\ket{0}_k + (-1)^{\hat{n}_k}\ket{1}_k\right)\ket{\Psi}\\
&= 2^{-L/2}  \prod_{k=0}^{L-1} \left(\sum_{s_k\in\{0,1\}}\ket{s_k}_k(-1)^{s_k\hat{n}_k}\right)\ket{\Psi}\\
&=2^{-L/2} \sum_{\mathbf{s}\in\{0,1\}^L}\left( \prod_{k=0}^{L-1} \ket{s_k}_k(-1)^{s_k\hat{n}_k}\right)\ket{\Psi},
\end{aligned}
\end{equation}
where $\ket{s_k}_k$ is the $k$-th ancillary qubit and $\hat{n}_k$ is the photon number operator acting only on the $k$-th pulse. After performing projection measurements on the ancillary qubits in the $Z$-basis, a specific measurement outcome, $\mathbf{s} = (s_0, s_1, \dots s_{L-1})$, corresponds to the final output of
\begin{equation}\label{eq:output}
\begin{aligned}
\ket{\Psi}_{\mathbf{s}} = \left( \prod_{k=0}^{L-1}(-1)^{s_k\hat{n}_k}\right)\ket{\Psi}.
\end{aligned}
\end{equation}

For instance, the state in Eq.~\eqref{eq:singlephoton} corresponds to the case when $\ket{\Psi}$ is a single photon state. The measurement outcomes of the $L$ ancillary qubits in the $\{\ket{0}, \ket{1}\}$ basis can be regarded as random numbers used to construct $L$ bits $\mathbf{s}$ in (Fig.~\ref{RRDPS:protocol}).

We suppose Alice measures the ancillary qubits after Bob announces $(i,j)$. To obtain $s_i\oplus s_j$, Alice performs a controlled-NOT gate (C-NOT) on the $i$-th and $j$-th ancillary qubits and measures the target in the Z basis. To define the phase error of the target qubit, we can measure it in the X basis. If the qubit is $\ket{+}$, no information is leaked to Eve. The phase error probability is denoted as the probability that the result is $\ket{-}$, which quantifies the leaked information of $s_i\oplus s_j$.

\section{Phase error estimation}\label{Sec:Securityanalysis}
The schematic for the RRDPS presented in figure~\ref{RRDPS:protocol} is shown in figure~\ref{Fig:RRDPS}\textbf{a}. To estimate how much sifted key information is leaked to Eve, one can consider an equivalent scenario, which is only applied in the security analysis, as shown in figure~\ref{Fig:RRDPS}\textbf{b}. In scenario \textbf{b}, Bob first generates a random number $r\in\{-L+1,\cdots,-1,1,\cdots,L-1\}$. Then, he measures the photon of the received signal and obtains a detection in the $i$th pulse. Bob calculates $j = i + r (\mathrm{mod}~L)$ with $i$ and $r$, and announces the values of $i$ and $j$. We consider Bob to be a black box with a quantum input and a classical output $(i, j)$ where Eve's interference of the quantum signal is considered. Since the input and output for the black boxes are identical under both scenarios, one can use scenario figure~\ref{Fig:RRDPS}\textbf{b} to estimate the phase error rate in scenario figure~\ref{Fig:RRDPS}\textbf{a}.

\begin{figure}[bht]
\centering \resizebox{12cm}{!}{\includegraphics[scale=1]{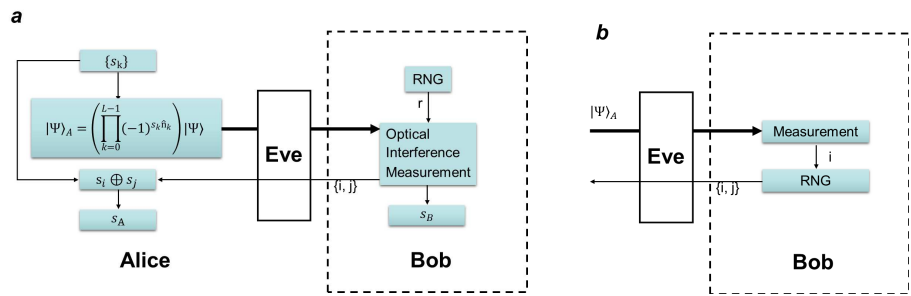}}
\caption{Two scenarios for the RRDPS protocol that are equivalent from Eve's viewpoint. \textbf{a}. The key distribution procedures are described in figure~\ref{RRDPS:protocol}. Bob receives a quantum state and announces $(i, j)$. \textbf{b}. Bob measures the location of the photon in the received signal to obtain $i$, and defines $j = i + r$ with randomly generated number $r\in\{-L+1,\cdots,-1,1,\cdots,L-1\}$.
} \label{Fig:RRDPS}
\end{figure}

In scenario $\mathbf{b}$, we imagine that Alice performs her measurement after Bob announces his outputs. To get the key bit $s_A = s_i \oplus s_j$, Alice simply applies a C-NOT to the $i$-th and $j$-th ancillary qubits, with the $i$-th qubit as the control and the $j$-th qubit as the target. After that, she measures the $j$-th qubit in the $Z$-basis and obtains a sifted key bit, $s_A$. To estimate the phase error rate, $e_{\textrm{ph}}$, one can simply measure the $j$-th qubit in the $X$-basis. If the $j$-th qubit is an eigenstate of the $X$-basis, the measurement outcome on the $Z$-basis is fully random that is, no information is leaked to Eve \cite{Koashi:Comp:09}. Hence, the phase error probability of measuring the $j$-th qubit is defined by the probability of finding it in the state $\ket{-}$.
As the C-NOT operation will not affect the $X$-eigenvalues of the $j$-th qubit, which are randomly chosen uniformly from all qubits except the $i$-th one by Bob, the phase error rate can be estimated by the probability of finding any except the $i$-th qubit in the $\ket{-}$ state. Notice that, in Eq.~\eqref{eq:State}, the probability of obtaining $+$ or $-$ is entirely determined by the number of photons contained in the $j$th pulse. The case of an odd (even) number of photons corresponds to outcome of $\ket{-}$ (resp.~$\ket{+}$). Therefore, according to Eq.~\eqref{eq:output}, the phase error rate can be estimated by the probability of finding an odd number of photons in a pulse.

According to Eq.~\eqref{eq:output}, the phase error rate can be upper-bounded by the probability of finding an odd number of photons appearing in a pulse. In the case where $\ket{\Psi}$ is an $n$-photon state, the maximum possible number of pulses wherein odd numbers of photons appear is $n$. In the SYK analysis \cite{sasaki2014practical}, the phase error rate is bounded by
\begin{equation}\label{original:eph}
\begin{aligned}
e_{\textrm{ph}}^{n}\le \frac{n}{L-1}.
\end{aligned}
\end{equation}

\section{GLLP analysis} \label{Sec:Key}
In practice, a phase-randomised weak coherent state photon source is widely used in QKD systems. In the RRDPS protocol, Alice prepares a phase-randomised coherent state pulse with intensity $L\mu$. According to the photon number channel model \cite{Lo:PRL:2005}, the state can be regarded as a statistical mixture of $n$-photon states,
\begin{equation}\label{eq:WCP}
\begin{aligned}
\rho = \sum_{n = 0}^\infty e^{-L\mu}\frac{(L\mu)^n}{n!}\ket{n}\bra{n}.
\end{aligned}
\end{equation}
Then, following the procedures presented in figure~\ref{RRDPS:protocol}, this strong pulse is split into $L$ identical small pulses through beam splitters and becomes the initial state $\ket{\Psi}$, which is encoded with key information according to Eq.~\eqref{eq:output}. Note that the intensity of each small pulse, $\mu$, is weak, but $L\mu$ can be large.

For each $n$-photon term in Eq.~\eqref{eq:WCP}, the phase error rates can be estimated by Eq.~\eqref{original:eph}. Denote the ratio of the key that needs to be sacrificed for privacy amplification by $H_{\mathrm{PA}}$; by extending the GLLP security analysis \cite{GLLP:2004}, the amount of key loss in privacy amplification is given by
\begin{equation}\label{Eq:GLLP}
\begin{aligned}
Q_{L\mu}H_{\mathrm{PA}} &= e^{-L\mu}\sum_{n=0}^\infty Y_n\frac{(L\mu)^n}{n!}H(e_{\textrm{ph}}^n),
\end{aligned}
\end{equation}
where $Q_{L\mu}=e^{-L\mu}\sum_{n = 0}^\infty Y_n{(L\mu)^n}/{n!}$ is the overall gain and $Y_n$ denotes the yield of the $n$-photon state.

Then, the final key rate, similar to Eq.~\eqref{keyrateN}, can be rewritten as
\begin{equation}\label{eq:extGLLP}
\begin{aligned}
L\cdot R&=Q_{L\mu}\left[1-H(e_{\textrm{bit}})-H_{\mathrm{PA}}\right],
\end{aligned}
\end{equation}
where $L\cdot R$ is the final key bit per $L$-pulse train. Since these trains contain $L$ pulses, the final key rate, $R$, should be normalised by $L$. In experiment, the overall gain $Q_{L\mu}$ is an observable, while $Y_{n}$ is generally an unknown parameter that can be manipulated by Eve. In the following, we show three different approaches to the estimation of $H_{\mathrm{PA}}$.

Let us start with the original SYK analysis \cite{sasaki2014practical}, where the phase error rate is estimated by Eq.~\eqref{original:eph}. One can set a threshold photon number $n_{th}$, over which the phase error rate is bounded by 1/2. Since the phase error rate $e_{\textrm{ph}}^{n}$ increases with the photon number $n$, one can consider the worst case scenario to be the case where the losses are all contributed from low photon numbers. That is, $Y_n = 1$ for $n>n_{\mathrm{th}}$. Also, for all the states with photon numbers less than $n_{th}$, one has $e_{\textrm{ph}}^{n}\le e_{\textrm{ph}}^{n_{th}}$. Thus, $H_{\mathrm{PA}}$ in Eq.~\eqref{Eq:GLLP} can be upper bounded by
\begin{equation}\label{nature:Heph}
\begin{aligned}
Q_{L\mu}H_{\mathrm{PA}}&\le \left(Q_{L\mu}-\sum_{n>n_{\textrm{th}}} \frac{(L\mu)^n}{n!}e^{-L\mu}\right)H(e_{\textrm{ph}}^{n_{\textrm{th}}})+ \sum_{n>n_{\textrm{th}}}\frac{(L\mu)^n}{n!}e^{-L\mu}H\left(\frac12\right),
\end{aligned}
\end{equation}
where $e_{\textrm{ph}}^{n_{\textrm{th}}}$ is bounded by Eq.~\eqref{original:eph}. In addition, one can optimise over the choice of $n_{\mathrm{th}}$ to minimize $H_{\mathrm{PA}}$ and hence maximize the final key rate $R$.

With the tagging technique developed in the GLLP security analysis, we can estimate each privacy-amplification term in Eq.~\eqref{Eq:GLLP} separately. According to Eq.~\eqref{original:eph}, the phase error rate increases with the photon number $n$. In the worst case scenario, we assume all the losses come from the low-photon number states (with $n<n_{\textrm{th}}$), whereas all of the high-photon number states (with $n>n_{\textrm{th}}$) pass through the channel transparently. Then, $H_{\mathrm{PA}}$ in Eq.~\eqref{eq:extGLLP} can be upper-bounded by
\begin{equation}\label{nodecoy}
\begin{aligned}
Q_{L\mu}H_{\mathrm{PA}}&\le \left(Q_{L\mu}-\sum_{n>n_{\textrm{th}}} \frac{(L\mu)^n}{n!}e^{-L\mu}\right) H(e_{\textrm{ph}}^{n_{\textrm{th}}}) + \sum_{n>n_{\textrm{th}}} \frac{(L\mu)^n}{n!}e^{-L\mu} H(e_{\mathrm{ph}}^n),
\end{aligned}
\end{equation}
where $e_{\textrm{ph}}^{n_{\textrm{th}}}$ and $e_{\mathrm{ph}}^n$ are bounded by Eq.~\eqref{original:eph}. Here, the threshold photon number, $n_{\textrm{th}}$, is the critical photon number such that the total, gain $Q_{L_\mu}$, can be obtained by contributions from the terms with $n\ge n_{\textrm{th}}$. In general, the value of $n_{\textrm{th}}$ calculated in Eq.~\eqref{nodecoy} is different from the optimal $n_{\textrm{th}}$ from the SYK analysis, Eq.~\eqref{nature:Heph}.

Although the yields $Y_n$ in Eq.~\eqref{Eq:GLLP} cannot be directly measured by experiments, we can use the decoy-state method, by which all the values of $Y_n$ can be accurately estimated with an infinite number of decoy states \cite{Lo:PRL:2005}. In the simulation, we simply use the case where Eve does not interfere with the yields,
\begin{equation}\label{Eq:Yn}
\begin{aligned}
Y_n&=1-(1-Y_0)(1-\eta)^n,
\end{aligned}
\end{equation}
where $\eta$ is the channel transmittance and $Y_0$ is the background count rate. That is, $Y_0$ denotes the count rate when Alice sends nothing ($n=0$).

\section{Simulation model and result}\label{Sec:Simulationmodel}
With the key rate formula for the RRDPS protocol given in Eq.~\eqref{eq:extGLLP}, we can compare the performances of the three different methods of estimating $H_{PA}$ namely, Eqs.~\eqref{nature:Heph}, \eqref{nodecoy} and \eqref{Eq:Yn}, by means of modelling a practical system \cite{XMA:PRA:2005}. The simulation model is presented in \ref{App:bitflip}, and the QKD experimental parameters are listed in Table \ref{table1}. In the simulation, we need to consider all the device imperfections such as misalignment, environmental noise and dark counts.

\begin{table}[htb]
\centering  %
\caption{Parameters from a typical QKD system \cite{GYS}. Here, $\eta_d$ is the detection efficiency, $\alpha$ is the channel loss, $e_d$ is the misalignment error rate, and $y_0$ is the background rate for each pulse. As there are $L$ pulses, the total background rate should thus be $Y_0=1-(1-y_0)^L\approx Ly_0$. We discuss the case where the total background rate is independent of $L$ in the Discussion section.}\label{table1}
\begin{tabular}{lcccc}
\hline
Experiment& $\eta_d$ & $e_d$ &$y_0$ & $\alpha$ \\
\hline
GYS \cite{GYS}& 4.5\% & 3.3\% &$1.7\times10^{-6}$ & 0.2 dB/km \\
\hline
\end{tabular}
\end{table}

The performances of the RRDPS protocol with different analytical methods: SYK analysis, new analysis (no-decoy) and new analysis (decoy) are shown in figure~\ref{eps32}. Here, we fix $L$ at $32$ and optimise $\mu$ to obtain the maximum transmission distance. As one can see from figure~\ref{eps32}, the improved analysis method enhances the performance, both in terms of the key rate and the maximum transmission distance. The simulation result indicates that the decoy-state method is useful for the RRDPS protocol.

\begin{figure}[hbt]
\centering \resizebox{12cm}{!}{\includegraphics{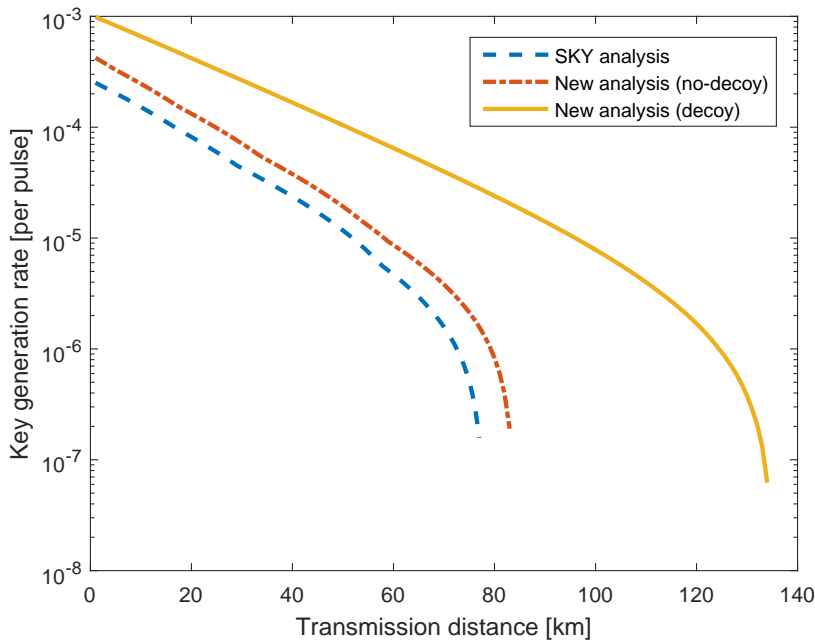}}
\caption{Final key rate for $L=32$ using a practical QKD model, with parameters listed in Table \ref{table1}.}
\label{eps32}
\end{figure}

In the conventional BB84 protocol, the decoy state is also utilized to increase the secure key generation rate and the transmission distance. Interestingly, the maximal secure distance of the asymptotic limit of the decoy state BB84 protocol (with infinite decoy states) is also around 140 km \cite{Lo:PRL:2005}, with the same set of experimental parameters. In the simulation, we compare the BB84 and RRDPS protocols. The result shows that the RRDPS protocol tolerates the misalignment error better (see figure~\ref{edrate}). Here, we compare the two protocols under two typical cases, for which transmission distances are, 50 km and 100 km, respectively. As shown in figure~\ref{edrate}, the final key rates of the RRDPS protocol are higher than those in the BB84 protocol when the misalignment error rate are greater than $7\%$. In the 50 km case, the RRDPS protocol can tolerate a misalignment error rate of more than $40\%$; in the 100 km case, secure key can be generated even if the misalignment error rate is equal to $25\%$ which is a hard upper bound of the BB84 protocol because of the intercept-and-resend attack \cite{Bennett:BB84:1984,PhysRevLett.92.217903}.

\begin{figure}[tbh]
    \centering \resizebox{12cm}{!}{\includegraphics{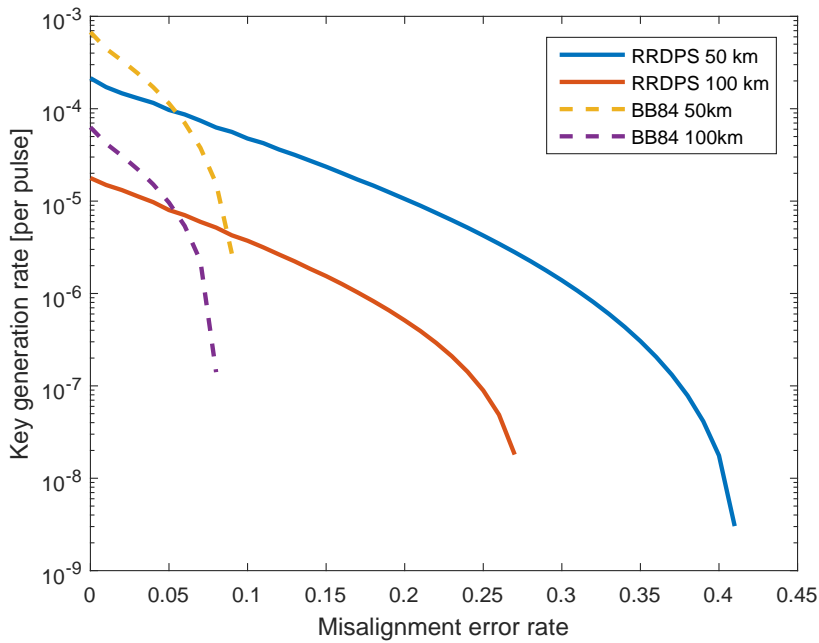}}
\caption{The optimized final key rate via the misalignment error rate.}
\label{edrate}
\end{figure}

We next briefly compare the RRDPS protocol with the measurement-device-independent QKD (MDIQKD) \cite{Lo:MDIQKD:2012,Stefano:MDIQKD:2012} protocol. The MDIQKD protocol has been demonstrated over 200 km \cite{Tittel:MDIQKDFielfTest:2013,Liu:MIQKDexp:2013,tang:experimental:2013,tang2014measurement} and in field test \cite{tang2014field}. While the MDIQKD protocol enjoys the advantage of being secure against any detection loopholes, the RRDPS protocol is able to tolerate higher error rate. In the short distances, similar to the BB84 protocol, the RRDPS protocol should yield a higher key rate than the MDIQKD protocol. We expect two protocols should find suitable applications in different practical scenarios.

\section{Discussion}\label{Sec:discussion}
In the original security analysis \cite{sasaki2014practical}, the signal going to Bob's detection box is assumed to be single photon states. Also, the detectors used by Bob are assumed to be single-photon detectors (or photon number resolving detectors). In practice, these requirements are challenging with current technology. Instead, normally coherent state sources and threshold detectors are used. Thus, there is a gap between the security analysis and the implementation. A similar problem also exists for other QKD schemes \cite{Ma2007Entangled}. The solution there is to apply the squashing model \cite{PhysRevLett.101.093601,PhysRevA.84.020303,PhysRevA.89.012325} to Bob's measurement. As a result, the signal Bob receives can be regarded as a qubit state in the security analysis. However, the squashing model cannot be directly applied here, since the single-photon state received by Bob is a qudit with a dimension of $2^L$. Thus, it is an interesting future project to work a squashing model for the general qudit case.

The upper bound of the phase error with $n$-photons given in Eq.~\eqref{original:eph} is only a rough estimation. An interesting future work is to find a tighter upper bound of the phase error rate and combine it with the GLLP tagging idea and the decoy-state method. In \ref{App:phase}, we discuss an ideal case where the input $n$-photons are considered to be independent. We show that the phase error estimation can be improved such that it becomes 1/2 only in the limit of infinite photon numbers while the original phase error becomes 1/2 when $n\ge L/2+1$. Although such an ideal scenario may become vulnerable in practice, the result may still shed light on a better upper bound to the phase error rate with multi-photons.

In practice, the parameter $L$ may not be chosen freely. When $L$ increases, the (relative) phase maintenance may becomes challenging. That is, it is reasonable to assume that the misalignment parameter grows with increasing $L$. Supposing that we ignore this practical issue for the moment and optimise the parameter $L$, our simulation result shows that with a large $L$ (optimal value around $10^4$ for the two no-decoy cases), three curves in figure~\ref{eps32} can reach a maximal secure distance of 140 km. In the decoy state case, the result is very stable under different values of $L$. In fact, with $L=32$, the performance is already very close to the optimal $L$ case. From this perspective, the decoy-state method makes the RRDPS protocol easier to implement in practice. Note that a practical decoy-state method based on our result is recently published \cite{Zhang:16}.

In the simulation, we assume that the total background count rate ($Ly_0$) in an $L$-pulse block would linearly increase with $L$. One can also consider a scenario where the background noise, $Y_0$, has a fixed value, independent of $L$. Under this assumption, as shown in \ref{App:distance}, we prove that the maximum transmission distance can infinitely increase, and that the optimal value of $L$ linearly increases with the inverse of the channel transmittance, $L\propto1/\eta$. This is not surprising, since the phase error rate approaches $0$ as $L$ increases, which allows the bit error rate (if it is independent of $L$) to grow arbitrarily close to 1/2.

Furthermore, we show in \ref{App:bitflip} that the RRDPS protocol can tolerate the misalignment error, $e_d$, up to $50\%$. Intuitively, this can be explained by the fact that the misalignment error ($e_d$) is independent of $L$, and it is similar to the case where the background noise, $Y_0$, is independent of $L$. Thus, we conjecture that the RRDPS is able to tolerate errors that are independent of $L$.

In this study, we make use a phase-randomised coherent state as input. In experiments, the continually phase-randomisation requires the phase uniformly distributed from 0 to $2\pi$, which is generally hard to implement. Instead, the discrete phase-randomisation can be applied to approximate exact phase-randomisation \cite{Discretephase2015}. We leave such an extension to future research.

\appendix

\section{Simulation model}\label{App:Simulation}
Here, we adopt a widely used simulation model for QKD \cite{XMA:PRA:2005}. Use $L\mu$ to denote the intensity of the source; $\eta$ to be the overall transmittance; $Y_n$ and $e_n$ to be the yield (the probability of obtaining a successful detection) and the error rate, respectively, with $n$ denoting the number of photons Alice sends. Without Eve's interference, $Y_n$ and $e_n$ are given by \cite{XMA:PRA:2005}
\begin{equation}\label{Yi:Qi}
\begin{aligned}
Y_n&=1-(1-Y_0)(1-\eta)^n,\\
e_nY_n&=e_0Y_0+e_d(1-Y_0)[1-(1-\eta)^n],\\
\end{aligned}
\end{equation}
where the value of $e_0$ is equal to $0.5$.
Then the overall gain and QBER are given by
\begin{equation}\label{Qi}
\begin{aligned}
Q_{L\mu}&=\sum Y_n\frac{(L\mu)^n}{n!}e^{-L\mu}\\
&=Y_0+(1-Y_0)(1-e^{-\eta L\mu}),\\
E_{L\mu}Q_{L\mu}&=\sum e_nY_n\frac{(L\mu)^n}{n!}e^{-L\mu}\\
&=e_0Y_0+e_d(1-Y_0)(1-e^{-\eta L\mu}).
\end{aligned}
\end{equation}
In simulation, we consider two different scenarios, an idealized one and a practical one. In the idealized case, we consider that the background noise, $Y_0=y_0$, is independent of $L$. In the practical condition, Bob is required to obtain $L$ detections and the background noise, $Y_0$, becomes $1-(1-y_0)^L$.

We show in \ref{App:distance} that the maximal transmission distances of the RRDPS protocol behave differently under each of these two scenarios. In the idealized case, we show that the maximal transmission distance of the RRDPS protocol can be infinite. In the practical case, we show that there exists a limit on the transmission distance (loss).

\section{Potential improvement for phase error rate estimation}\label{App:phase}
In this section, we consider phase error estimation with an $n$-photon state as a comparison to the estimation given in Eq.~\eqref{original:eph}. Here, we consider an ideal scenario that the $n$ photons are independent. Note that the $n$ photons are indeed independent when Alice prepares the state. Thus, the ideal scenario considered here only assumes that the quantum channel preserves this independency, for example, the beam splitting channel model.


In such a case, we can consider that each photon independently appears in each pulse with an equal probability of $p=1/L$. One can imagine that Alice first prepares an $n$-photon state and allows it to pass through many beam splitters to form an $L$-pulse sequence. We refer to Refs.~\cite{frohlich2015quantum,takesue2015experimental} for the details of experimental implementations. When considering the case where Eve's operation in the quantum channel does not change the photon-number statistics, the phase error rate estimation can be improved over the original one \cite{sasaki2014practical},
\begin{equation}\label{eq:Neph}
\begin{aligned}
e_{\textrm{ph}}^n &= \sum_{k\in \mathrm{odd}} {\binom n k} p^k(1-p)^{n-k} \\
&= \frac{1 - \left(1 - 2p\right)^{n}}{2}. \\
\end{aligned}
\end{equation}
The key point here is that in the RRDPS protocol, the phase error rate of each pulse is determined by the preparation of quantum state, but not by Eve's interaction. Thus, Alice and Bob do not need to accept the worst case scenario; instead, they can accurately derive the phase error rate in the state-preparation stage. This is the essential reason why the phase error rate in the RRDPS protocol is independent of the bit error rate.

Before we apply the new phase error estimation method to the QKD scheme, let us first compare the SYK result in Eq.~\eqref{original:eph} with the new one in Eq.~\eqref{eq:Neph} in figure~\ref{fig:EL32}. One can see that the improved method does give a tighter bound on the phase error rate, $e_{\textrm{ph}}^n$, for an $n$-photon state source. We expect that the key rate will be improved by employing the improved scheme, and this is confirmed in later simulations.

\begin{figure}[tbh]
\centering \resizebox{12cm}{!}{\includegraphics{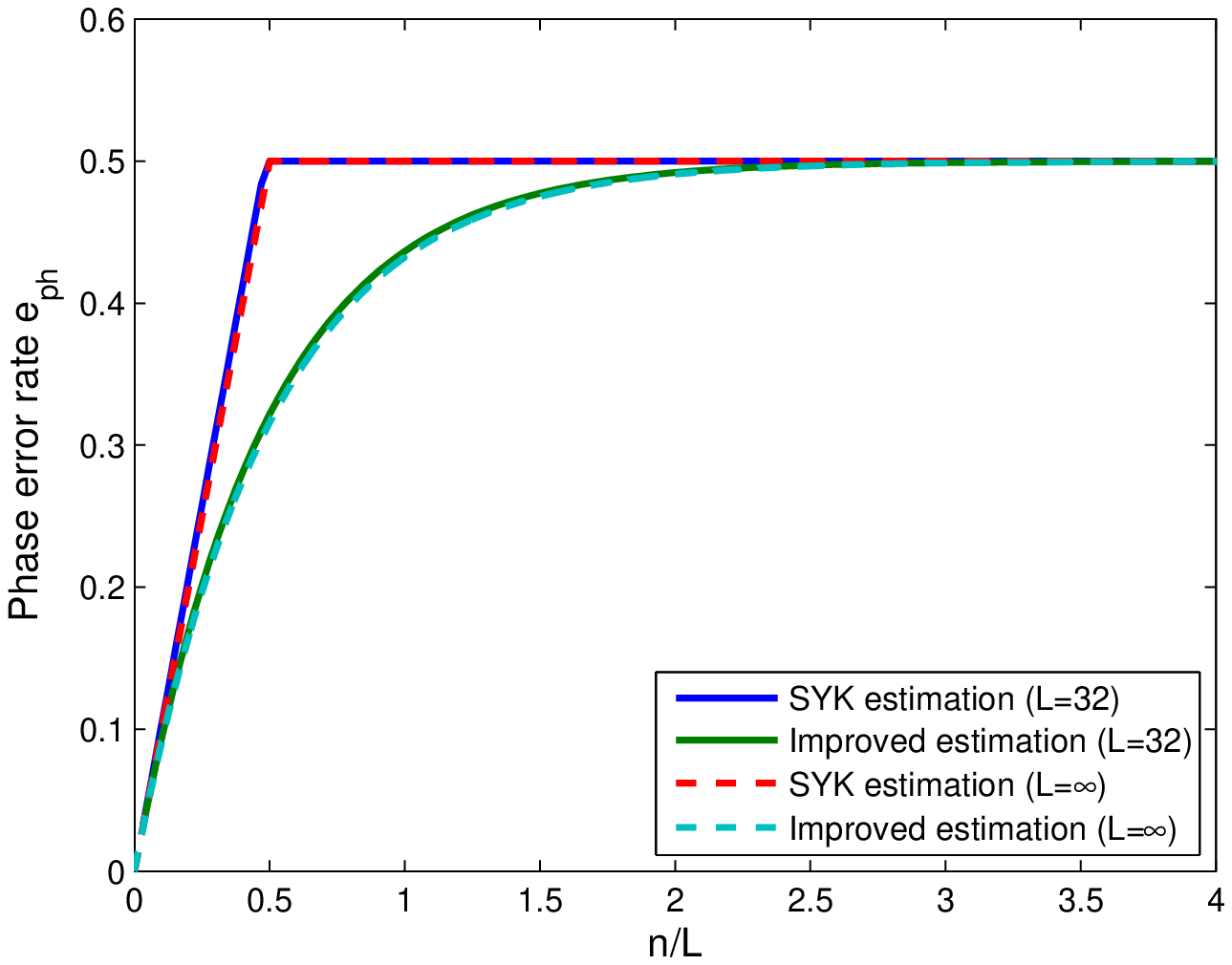}}
\caption{Comparison between two different estimates of $e_{\textrm{ph}}^n$ with $L=32$ and $L=\infty$, in the case where Eve does not change the photon-number statistics. The SYK analysis method is given in Eq.~\eqref{original:eph}. The improved method is given in Eq.~\eqref{eq:Neph}.} \label{fig:EL32}
\end{figure}

As shown in Fig.~\ref{fig:EL32}, when the total photon number $n$ of the $L$-pulse quantum signal  increases beyond the value of $L$, the phase error rate, $e_{\textrm{ph}}^n$, exponentially approaches to 1/2 quickly, that is,
\begin{equation}\label{eq:leeph}
\begin{aligned}
e_{\textrm{ph}}^n \approx \frac{1}{2}-e^{-2n/L},
\end{aligned}
\end{equation}
where the ratio $n/L$ can be interpreted as the mean photon number of each pulse. On the other hand, when $n$ is much smaller than $L$, $e_{\textrm{ph}}^n$ can be approximated as
\begin{equation}\label{eq:NephApp}
\begin{aligned}
\frac{1}{e_{\textrm{ph}}^n} &= \frac{2}{1 - \left(1 - \frac{2}{L}\right)^{n}} \\
&\approx \frac{2}{1-\left(1- \frac{2n}{L}+\frac{2n^2}{L^2}\right)}-\frac{L}{n}+\frac{L}{n}\\
&\approx\frac{L}{n}\left( \frac{1}{1-\frac{n}{L}}-1\right)+\frac{L}{n}\\
&\approx 1+\frac{L}{n}.\\
\end{aligned}
\end{equation}
It is not hard to see that the phase error decreases along with the mean photon number of each pulse. In fact, in the entire regime of $n$ and $L$, the phase error rate, $e_{\textrm{ph}}^n$, mainly depends on the average photon number per pulse, $n/L$, as seen in Eq.~\eqref{eq:NephApp}. In the meantime, we can see from figure \ref{fig:epsephcost} that the new estimation method defined in Eq.~\eqref{eq:Neph} is always better than the original SYK method defined in Eq.~\eqref{original:eph}. Although the ideal case considered here is not the worst case scenario in practice, the improvement here indicates a better potential theoretical bound to the phase error estimation with multi-photon states.

\begin{figure}[tbh]
\centering \resizebox{12cm}{!}{\includegraphics{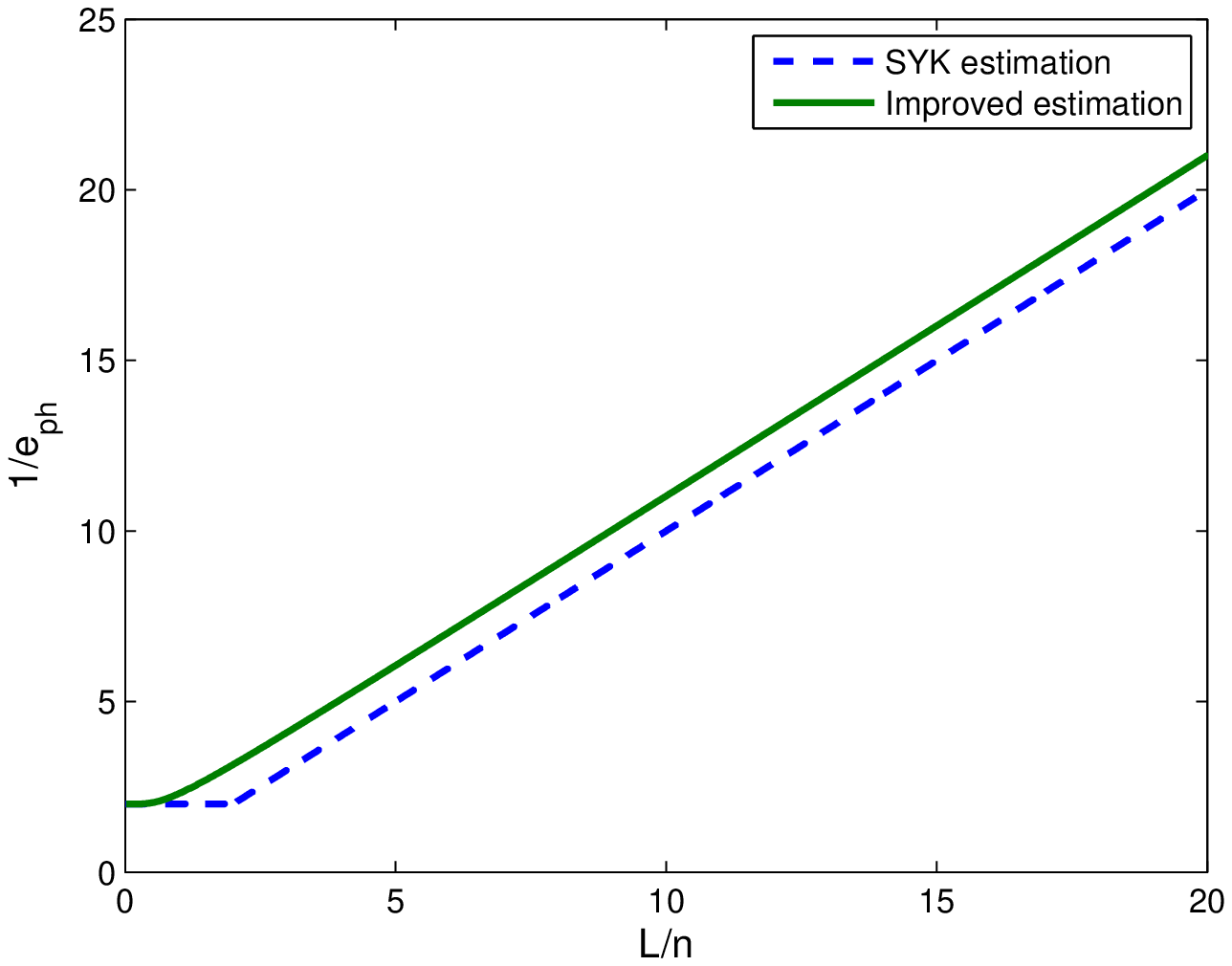}}
\caption{Comparison of two different estimates of $1/e_{\textrm{ph}}^n$ as a function $L/n$. The SYK analysis method is given in Eq.~\eqref{original:eph}; the new method is given in Eq.~\eqref{eq:Neph}. When the SYK analysis method is applied, the optimized value of the mean photon number is around $0.05$. In this case, the value of $L/n$ is around $20$. } \label{fig:epsephcost}
\end{figure}

\section{Maximal transmission distance}\label{App:distance}
To calculate the maximal transmission loss, we consider the asymptotic case where the values of $L$ and $L\mu$ are very large. Since the total photon number of the state prepared by Alice follows a Poisson distribution, the photon number can be well-approximated by $L\mu$. In this case, the cost of privacy amplification is close to a fixed value. A secure key can be generated in the case where the final key rate, $R$, in Eq.~\eqref{keyrateN} is bigger than $0$:
\begin{equation}\label{Rge0}
\begin{aligned}
&1-H(e_{\textrm{bit}})-H(e_{\textrm{ph}})\ge0.\\
\end{aligned}
\end{equation}
According to Eqs.~\eqref{original:eph} and~\eqref{Rge0}, the threshold value of the bit-flip error rate, $c$, is
\begin{equation}\label{eq:C}
\begin{aligned}
c&=H^{-1}\left[1-H\left( e_{\textrm{ph}}\right)\right]=H^{-1}\left[1-H\left( \frac{L\mu}{L-1}\right)\right]\approx H^{-1}\left[1-H\left( \mu\right)\right].\\
\end{aligned}
\end{equation}
Based on the simulation model in Eq.~\eqref{Qi}, the bit-flip error rate is given by
\begin{equation}\label{ebitqi}
\begin{aligned}
e_{\textrm{bit}}=\frac{E_{L\mu}Q_{L\mu}}{Q_{L\mu}}=e_d+\frac{(0.5-e_d)Y_0}{Y_0+(1-Y_0)(1-e^{-\eta L\mu})},\\
\end{aligned}
\end{equation}
which is a decreasing function of the overall transmittance $\eta$. Since the bit error rate $e_{\textrm{bit}}$ is upper-bounded by $c$, as given in Eq.~\eqref{eq:C}, the minimal overall transmittance $\eta_{\textrm{min}}$, in the case where Alice and Bob can communicate securely, can be calculated accordingly. Considering $e_{\textrm{bit}}\le c$, Eq.~\eqref{ebitqi} can be rewritten as
\begin{equation}\label{Min:eta1}
\begin{aligned}
1-e^{-\eta_{\textrm{min}} L\mu}= \left(\frac{0.5-e_d}{c-e_d}-1\right)\frac{Y_0}{1-Y_0}.\\
\end{aligned}
\end{equation}
Suppose that $\eta_{\textrm{min}}$ is small, the term $1-e^{-\eta_{\textrm{min}} L\mu}$ can be well-approximated by $\eta_{\textrm{min}} L\mu$. Then, the minimized $\eta_{\textrm{min}}$ can be approximated by
\begin{equation}\label{Min:eta}
\begin{aligned}
\eta_{\textrm{min}}&\approx\frac{1}{L}\left[\frac{1}{\mu}\left(\frac{0.5-e_d}{c-e_d}-1\right)\right]\frac{Y_0}{1-Y_0}.\\
\end{aligned}
\end{equation}

Notice that, the relationship between the transmission loss $Tl$ (dB) and the overall transmittance $\eta$ is defined by
\begin{equation}\label{loss:eta}
Tl = -10\log_{10}{\eta},
\end{equation}
and the relationship between the transmission distance $D$ (km) and the overall transmittance $\eta$ is
\begin{equation}\label{Dis:eta}
D = \frac{Tl}{\alpha} =-50\log_{10}{\eta},
\end{equation}
where the channel loss $\alpha$ is 0.2 dB/km, as we adopted in Table \ref{table1}.
In general, the transmission distance $D$ increases as the overall transmittance $\eta$ decreases.

\subsection{$L$-independent $Y_0$}\label{Sec:Idealassumption}
In an idealized case, where the background noise $Y_0=y_0$ is independent of $L$, secure transmission loss can be arbitrarily large with increasing $L$. That is, a secure key can be transmitted through arbitrarily large distance. Under this condition, $L\eta$, only depends on $\mu$, as shown Eq.~\eqref{Min:eta}. We can therefore optimise the parameter $\mu$ to minimize $L\eta$, which is found to be around $0.06$ when using the experimental parameters in Table \ref{table1}. Under this optimal $\mu$, the overall transmittance is a linear function of $1/L$, which can be infinitely small if $L$ is sufficiently large.

We can also estimate the final key rate in the case that is very close to the maximal transmission distance. For instance, we consider the regime where the transmission distance is $0.1$ km less than the maximal distance. In this regime, the optimal $L\eta$ can be considered to have a fixed value. Combined with the optimal value of $\mu$, the parameter $L\eta\mu$ is a fixed value. According to Eqs.~\eqref{Qi} and ~\eqref{ebitqi}, $Q_{L\mu}$, $e_{\textrm{bit}}$, and $E_{L\mu}Q_{L\mu}$, determined by $L\eta\mu$, are constants. The phase error rate $e_{\textrm{ph}}$ is determined by the parameter $\mu$. Thus, we can see that the right-hand side of the Eq.~\eqref{eq:extGLLP} is a constant, and the final key rate $R$ is a linear function of $1/L$ (or $\eta$).

\subsection{$L$-dependent $Y_0$ }\label{App:Practical}
Under a practical condition, the total background rate $Y_0$ also depends on $L$. Suppose the state Alice that prepared is a vacuum, the probability that Bob still obtains a successful detection in each pulse is a nonzero value, $y_0$, due to the background noise. Since there are $L$ pulses, the total background contribution $Y_0$ is defined by the probability of a successful detection event with the vacuum input, which can be given by $1-(1-y_0)^L$.

From Eqs.~\eqref{Min:eta} and ~\eqref{eq:upperbounded}, the overall transmittance is given by
\begin{equation}\label{p:Min:eta}
\begin{aligned}
\eta&\approx\frac{1}{L}\left[\frac{1}{\mu}\left(\frac{0.5-e_d}{c-e_d}-1\right)\right]\frac{1-(1-y_0)^L}{(1-y_0)^L}\\
&\ge\left[\frac{1}{\mu}\left(\frac{0.5-e_d}{c-e_d}-1\right)\right]y_0,
\end{aligned}
\end{equation}
where the second step can be derived by
\begin{equation}\label{eq:upperbounded}
\begin{aligned}
\frac{Y_0}{1-Y_0}&=\frac{1-(1-y_0)^L}{(1-y_0)^L}\\
&\ge(\frac{1}{1-y_0})^L-1\\
&\ge(1+y_0)^L-1\\
&\ge Ly_0.\\
\end{aligned}
\end{equation}

For any reasonable $\mu$, it can be concluded from Eq.~\eqref{p:Min:eta} that the lower bound of the overall
transmittance is a fixed value independent of $L$. Therefore, the transmission distance cannot reach infinity
under this practical condition.
Note that the bound of Eq.~\eqref{p:Min:eta} is not tight.

\section{Tolerable bit error rate}\label{App:bitflip}
In the main context, we have shown that the RRDPS protocol can tolerate a bit-flip error rate $e_{\mathrm{bit}}$ close to $0.5$ when the phase error rate $e_{\mathrm{ph}}$ tends to $0$. Here we give a simulation example to show that, under a practical condition, the RRDPS protocol can generate a secure key when $e_{\mathrm{bit}}=0.4923$. The result is shown in Table \ref{table2}. One can see that the RRDPS protocol can tolerate the bit-flip error well.
\begin{table}[htb]
\centering  %
\caption{RRDPS with a bit error rate close to $0.5$. The experimental parameters are listed in Table \ref{table1}, except $\eta_d=90\%$ and $Y_0=1-(1-y_0)^L$ (without approximation). Here, we employ our new analysis method with decoy states.}\label{table2}
\begin{tabular}{lcccccc}
\hline
Distance & $L$ &$L\mu$  &$e_d$&$e_{\mathrm{bit}}$& R\\
\hline
1 km & $220000$ &$0.77$  &$0.485$ &$0.4923$ &$ 2.265\times10^{-10}$\\
\hline
\end{tabular}

\end{table}


\section*{Acknowledgements.}
We thank Q.~Zhao and J.~Ma for helpful discussions. This study was supported by the National Natural Science Foundation of China Grants No.~11674193.

\section*{Reference}
\bibliographystyle{iopart-num}
\bibliography{bibRRDPS}
\end{document}